# Attosecond high-harmonic interferometry probes orbital- and band-dependent dipole phase in magnesium oxide

Nataliia Kuzkova[1,2][†], Pieter J. van Essen[1], Brian de Keijzer[1], Rui E. F. Silva[3,4], Álvaro Jiménez Galán[3,4], Peter M. Kraus[1,2][*]

[1]Advanced Research Center for Nanolithography (ARCNL), Science Park 106, 1098 XG Amsterdam, The Netherlands
[2]Department of Physics and Astronomy, and LaserLaB, Vrije Universiteit (VU), De Boelelaan 1105, 1081 HV Amsterdam, The Netherlands
[3]Instituto de Ciencia de Materiales de Madrid, Consejo Superior de Investigaciones Científicas (ICMM-CSIC), Sor Juana Inés de la Cruz 3, 28049 Madrid, Spain
[4]Max-Born-Institute for Nonlinear Optics and Short Pulse Spectroscopy (MBI), Max-Born-Strasse 2A, D-12489 Berlin, Germany

To whom correspondence should be addressed; E-mails: [†]n.kuzkova@arcnl.nl, [*]p.kraus@arcnl.nl.

**Control over the spatial coherence, wavefront, and focusability of emitted light relies on understanding the intrinsic phase of the emission process, and vice versa, measuring phase can reveal insights about microscopic generation mechanisms. A thorough understanding of the origin of the dipole phase in solid-state high-harmonic generation is currently missing. Here, by employing attosecond interferometry with phase-locked XUV pulses, we directly assess the intensity- and frequency-dependent dipole phases in magnesium oxide (MgO) in solid-state HHG. We also quantify a nonlinear phase shift of the fundamental and disentangle its contribution from the dipole phase. Theoretical models (analytical, two-band, and**

**full-band numerical simulations) support our results. The analytical approach aids future solid-state HHG experiments and simulations, while the full numerical model details orbital- and band-resolved current contributions to the dipole phase. Our research delivers the first combined quantitative measurement and rigorous theoretical description of the harmonic emission phase in solids.**

# Introduction

High-harmonic generation (HHG) from solids under strong laser fields presents a compelling frontier in nonlinear optics (*1–3*), unlocking new possibilities for ultrafast spectroscopy and attosecond science. Recent advancements in this field highlight the potential of solid-state HHG as a spectroscopy (*4–6*) and optically controlled super-resolution microscopy technique (*7, 8*), as well as a compact source of extreme ultraviolet (XUV) radiation, characterized by a high degree of temporal and spatial coherence. This coherence is essential for innovative applications such as lensless diffractive imaging and precision metrology at the nanoscale (*9, 10*), fueling the pursuit of all-solid-state HHG sources capable of generating coherent and focused XUV light. Achieving efficient HHG-based sources and optimal HHG beam focusability (*11–13*) requires careful consideration of the spatial coherence of emitted harmonics, which is fundamentally governed by the phase and divergence of the resulting XUV wavefront. The phase of high-harmonic emission, determined by the dipole phase, strongly depends on the intensity and frequency of the fundamental laser field (*14, 15*). As electrons in the atoms or molecules of the generating HHG medium interact with the intense laser field, they traverse various electronic quantum paths, ultimately recombining with their parent ions to emit XUV photons at harmonics of the driving frequency. In gases, this process is well understood through a semi-classical framework (*16–19*), where two energetically-degenerate pathways, known as the short and long electron trajectories, play a pivotal role in determining the dipole phase of emitted harmonics. There, the characteristics of the driving laser beam – such as wavelength, peak intensity, polarization state, and temporal profile – primarily shape the dipole phase, rather than the specific atomic species being used (*11, 20–24*). In solid-state HHG, however, the behavior is distinct from that in gases, as solids possess a well-defined electronic band structure that significantly impacts electron dynamics in the presence of strong laser fields (*25–27*). The microscopic response of HHG in solids showed that the laser field-induced transitions of electrons between valence and conduction bands introduce two principal contributions to the dipole phase: interband and intraband dynamics (*28–30*). Interband dynamics involve polarization effects that drive electronic excitation and electron-hole recombination, while intraband dynamics pertain to the oscillatory motion of tunnel-ionized electrons within a single band. Both contributions interact intricately with the laser field parameters and the material's electronic properties, leading to multifaceted phase characteristics in solid-state HHG responses.

Despite its importance, only a few recent experimental studies have focused on examining the dipole phase in solid-state HHG (*31, 32*). While linear relationships between the phase of solid-state HHG and intensity—similar to gas-phase HHG—have been established (*31*), a clear link to a microscopic model via theory beyond few-level simulations is missing. In light of this, the present study aims to close this knowledge gap by investigating the dipole phase in the solid-state HHG, bolstered by robust theoretical calculations. The complexities inherent to measuring and manipulating the dipole phase in solid-state HHG pose substantial challenges, demanding meticulously designed experiments and sophisticated theoretical modeling. However, recent advancements in XUV interferometry techniques have advanced our ability to probe HHG phenomena at the atomic level, both in gases and solids (*31–36*). XUV interferometry offers a unique and powerful approach for examining the dipole phase and its contribution to high-harmonic emission. By interfering two XUV pulses, the phase of the emitted harmonics can be directly quantified and obtained. Nevertheless, conducting XUV interferometry experiments often comes with challenges related to achieving and maintaining high stability and intrinsic interferometric delay precision at the sub-wavelength level, particularly in terms of phase control. The widely used XUV interferometers, such as those utilizing Wollaston prisms (*37*), curved two-segment mirrors (*36*), Mach-Zehnder (*38*) or Michelson-type (*39*) designs, often experience changes in the different optical paths due to misalignments, thermal expansion, or mechanical instabilities, resulting in phase variations and reduced stability. Thus, these interferometers require active stabilization of optical components and feedback loops to achieve fine control over the phase (*40, 41*).

In this work, we utilize an ultrastable birefringent common-path interferometer that incorporates two delayed, phase-locked collinear replicas of an input ultrashort near-infrared (NIR) laser pulse to generate two well-controlled XUV sources with attosecond precision. The interferometer design draws inspiration from the Translating-Wedge-Based Identical Pulses eN-coding System (TWINS) (*42*) and has been successfully implemented in multiple spectral regions (*42–46*). Here, we leverage XUV interferometry to assess the dipole phase of high-harmonics generated from bulk magnesium oxide (MgO) crystalline solid. We resolve the harmonic phase shifts by measuring the corresponding interference fringes in the far field, varying the relative peak intensities of two spatially separated fundamental 800 nm beams focused on the MgO target. This methodology allows us to directly probe the intensity- and frequency-dependent relative dipole phase of high-harmonics in the solid. Furthermore, through the dipole phase measurements, we reveal that our interferometric setup is sensitive to the nonlinear phase shift, namely the B-integral, accumulated as the intense laser beam travels through a nonlinear medium. We find that the contribution from this intrinsic parameter, reflecting the degree to which laser intensity modifies the refractive index of the material, can be individually identified and subsequently subtracted from the measured harmonic emission phase while maintaining the same experimental conditions. To the best of our knowledge, this marks the first quantitative measurement of the dipole phase and nonlinear effects occurring in real-time in solid-state HHG, achieved using an ultrastable XUV interferometer. To evaluate the stability of our interferometer and guarantee the accurate synchronization of the generated harmonics with the fundamental frequency, we perform temporal XUV interferometric measurements, which involve the interference of two time-delayed coherent XUV fields. By controlling their relative time delay with attosecond-level precision, we can accurately determine the time-dependent phase delays of individual harmonics within one optical cycle of an 800 nm driver. We implement several theoretical models and simulations to assist with the analysis of the intensity-dependent experimental data and clarify the physical origins of the harmonic dipole phase in bulk MgO. First, an analytical semi-classical HHG model is applied to elucidate the similarities and differences in harmonic dipole phase between gases and solids, given that the interband mechanism in solids exhibits a recollision-like behavior akin to that of gaseous atoms or molecules. Next, numerical simulations based on the semiconductor Bloch equations (SBE) are performed for the two-band system, incorporating the material's electronic band structure and the interplay between interband and intraband contributions to the dipole phase. Finally, we implement an orbital-based framework in the Wannier basis for the multi-band system using numerical SBE calculations, providing a clear picture of how different orbitals contribute to the overall harmonic dipole phase and shedding light on the underlying physics of the system.

## Results

**Spectrally-resolved XUV interferometry.** The spectrally-resolved XUV interferometric experiment in bulk MgO is depicted in Fig. 1A. An input femtosecond NIR pulse with the intensity $I_0$ is transformed into two NIR driving pulse replicas by a birefringent common-path interferometer, which are subsequently focused into a solid HHG medium to generate XUV pulses while maintaining a vertical separation between their focal points on the target. Following the generation of two spatially separated XUV sources from these foci, the emitted light is spectrally dispersed in the horizontal direction and spatially diverged vertically, yielding a spatio-spectral image with interference horizontal fringes of high-harmonic emissions captured in the far-field. The relative dipole phase of high-harmonics, $\Delta\varphi_q$, is assessed by varying the relative peak intensities of the two pulse replicas ($I_1$ and $I_2$) and measuring the intensity-induced fringe shifts ($\varphi(I_1)$ and $\varphi(I_2)$) of the individual harmonics $q$. The interferometer's reliability and precise synchronization of the produced harmonics with the fundamental frequency are evaluated by measuring the relative phase delay of high-harmonics as a function of time ($\Delta\tau$). This process entails examining the interference fringe shifts between two coherent XUV fields that are time-delayed while ensuring that the intensity distribution remains equal across the NIR foci ($I_1 = I_2$).

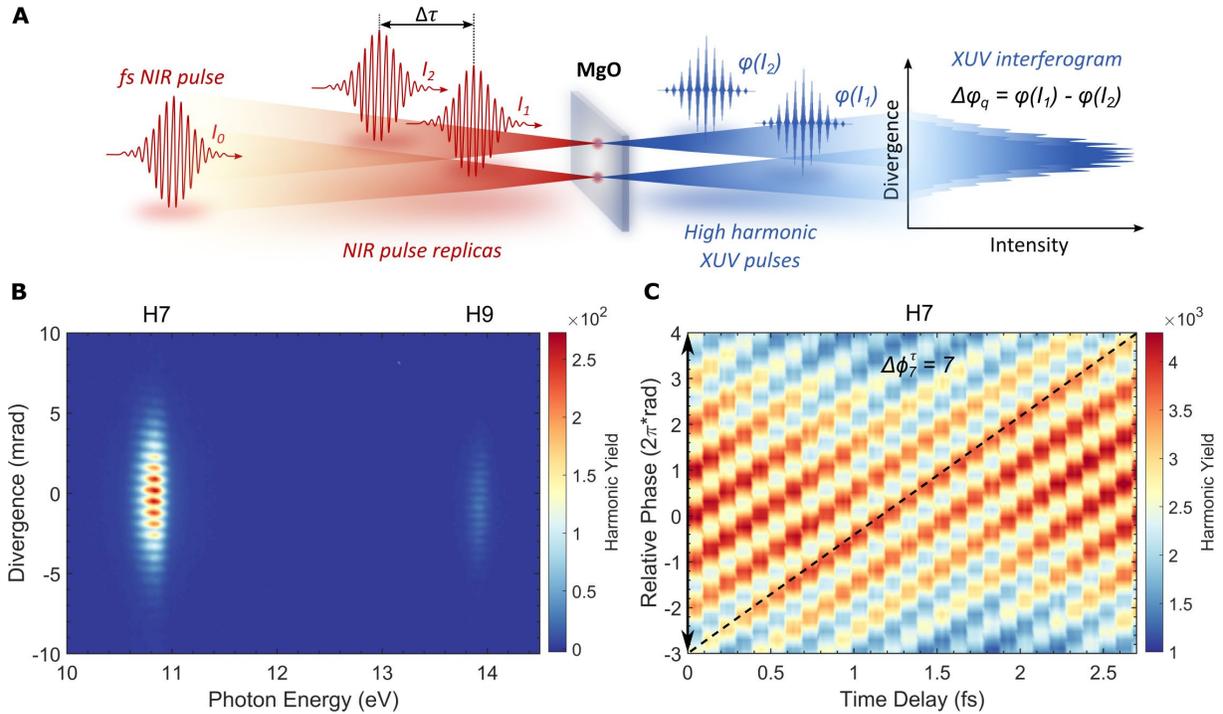

**Fig. 1. Spectrally-resolved XUV interferometry of MgO solid. (A)** Concept of the XUV interferometric experiment for measuring the phase of high-harmonic emission in solids. **(B)** The far-field XUV interferograms obtained from MgO, presented with the intensities of the NIR foci set to be identical (6 TW cm−2). The harmonic orders (H7 and H9) refer to the odd multiples of the 800 nm driver. **(C)** Time-dependent relative phase shift, $\Delta\varphi_q^\tau$, in MgO for harmonic order $q = 7$, as obtained from the temporal XUV interferograms. The relative fringe

shift of H7 is shown with a dashed line. The extracted $\Delta\varphi_7^\tau$ value over an optical cycle of 2.7 fs of the 800 nm laser field is indicated by a black arrow and is depicted in the figure.

Fig. 1B exemplary shows the far-field XUV interferograms obtained from the bulk MgO with both driving NIR beams having matched intensities at their spatio-temporal peak (6 TW cm$^{-2}$ at each focus). The odd-order harmonics such as harmonic 7 (H7, 10.8 eV) and harmonic 9 (H9, 13.9 eV) were generated with an 800 nm laser driver following the experimental conditions outlined in the Materials and Methods. The fringe patterns observed in the far-field interferograms on the detection plane, which are proportional to the separation distance of the NIR foci, were chosen to effectively resolve phase shifts in high-harmonics while avoiding oscillations at the fundamental frequency. Fig. 1C shows the 2D colour map representing the time-dependent relative phase shift of H7 in MgO, recorded as a function of the time delay between the NIR pulse replicas. The $\Delta\tau$ was varied by translating one of the birefringent wedges of the interferometer transversely over a $\Delta x = 30$ $\mu$m range with a step size of 0.2 $\mu$m corresponding to a scanned $\Delta\tau = 2.96$ fs range and a minimum step motion of 20 as. This was sufficient to follow one optical cycle of the 800 nm driving wavelength, $\tau_{800} \approx 2.7$ fs, with high spectral resolution. As a result, the time-dependent relative phase delays of the generated harmonics were defined with a sub-20-as accuracy, providing detailed insights into their temporal evolution. By measuring the displacement of the fringes over time for H7 and H9, we validated the robustness of our common-path interferometer and the precise synchronization of harmonics with the NIR driver, ensuring sub-cycle phase-locking. Since the two foci are identical, the dipole phase term (which relies on the driving field intensity) is the same for both, and therefore, it does not impact the temporal harmonic phase difference (*18*). In Fig. 1C, the relative phase shift of H7 over the $\tau_{800}$ range is fitted with a linear function (dashed line). The time-dependent relative phase of the harmonic field is determined within a 2.7 fs range (black arrow). Here, the maxima and minima of the fringe pattern indicate a $(2\pi) \cdot rad$ phase shift that represents the distance between two peaks and their relative displacement. The time delay affects the phase of the generated harmonics in a linear manner, expressed as (33): $\Delta\varphi_q^\tau = \tau_{800} \cdot q$, where $\Delta\varphi_q^\tau$ represents the relative phase shift associated with the $q$-th harmonic order. Analysis of the experimental data, as illustrated in Fig.1C, shows that the extracted phase shift for the H7 is $\Delta\varphi_q^\tau = 7\,(2\pi) \cdot rad$, which corresponds exactly to the harmonic order $q = 7$. Furthermore, temporal XUV interferometric results for the H9 corroborate that this linear dependence holds for higher harmonic orders (fig. S4). These findings underscore the exceptional resolution and sensitivity to the harmonic phase, achieved through the XUV interferometric approach employed here. Moreover, the absence of any fringe oscillation component corresponding to the periodicity of the fundamental (i.e., with a periodicity of 2.7 fs) indicates that the two HHG sources are well separated in the near field and do not overlap, which would otherwise obscure the subsequent extraction of the dipole phase.

**Harmonic relative dipole phase and nonlinear optical effects.** The intensity-induced relative dipole phase of high-harmonics in the MgO was investigated by changing the relative peak intensities between the NIR foci. The intensities of the NIR beams, $I_1$ and $I_2$, were varied over $\Delta I/I_0 = \pm 1$ range by rotating a half-wave plate (HWP) positioned in front of the interferometer (fig. S1), where $I_0$ is the sum of the input intensities $(I_1 + I_2)$ and $\Delta I$ is their difference $(I_1 - I_2)$. A value of $\pm 1$ means that all the intensity is placed into a single beam, resulting in no interference, while a value of 0 represents equal intensity between the two focal points. The HWP was scanned over a rotational angle, $\theta$, ranging from 0° (0 rad) to 45° ($\pi/4$ rad), with an increment of 0.1°($\pi/1800$ rad). The far-field XUV interferograms of the spectrally dispersed high-harmonics were recorded as a function of the $\theta$ position associated with the $\Delta I/I_0$. During the experiments, the distinct interference fringes were observed within the range of $\Delta I/I_0 = \pm 0.3$ (0.6 total varied intensity range), corresponding to a change from 3 to 9 TW cm$^{-2}$ in peak intensities (fig. S5). When the intensity at both focal points was equal, the peak intensity at each focus corresponded to 6 TW cm$^{-2}$. Thus, this specified range was selected to analyze the intensity-induced relative fringe shifts throughout all measurements.

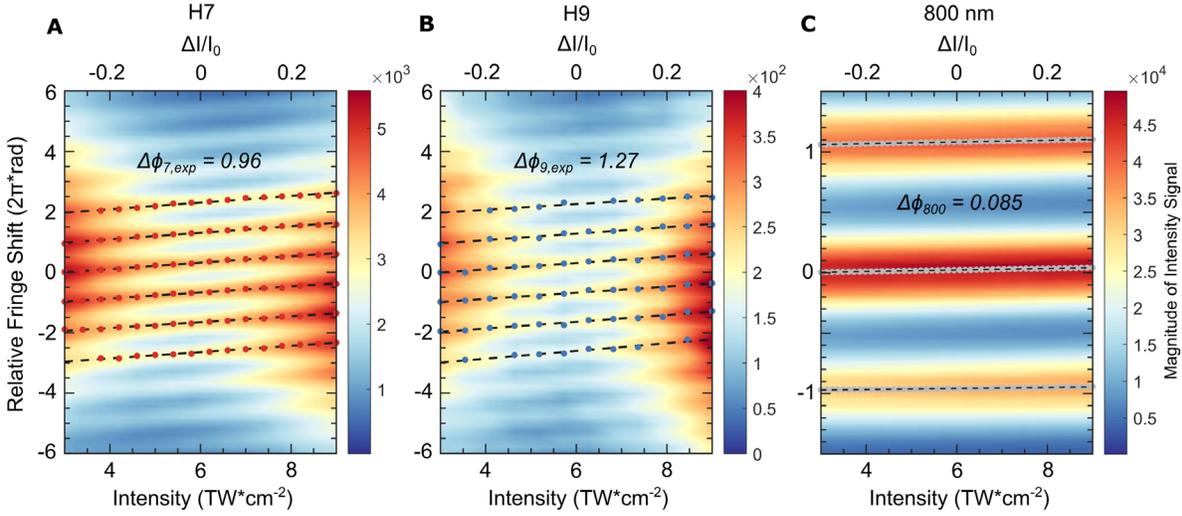

**Fig. 2. Intensity-induced relative fringe shifts for the harmonic and fundamental 800 nm emissions from MgO.** The relative fringe shifts, $\Delta\varphi_{q,exp}$, in MgO for H7 **(A)** and H9 **(B)** recorded as a function of varying relative peak intensities of the NIR foci within $\Delta I/I_0 = \pm 0.3$ range, corresponding to 3–9 TW cm$^{-2}$. **(C)** The nonlinear phase shift or B-integral, $\Delta\varphi_{800}$, which was accumulated in MgO sample at the fundamental 800 nm driver across the same measured intensity range. The $\Delta\varphi_{q,exp}$ and $\Delta\varphi_{800}$ values, depicted in the figures, were determined for the harmonic 7 (red dots), harmonic 9 (blue dots), and fundamental (gray dots) signals through linear fits to the fringe shifts (dashed lines) and the corresponding y-axis intercepts. The $\Delta\varphi_{q,exp}$ values, converted into radians after accounting for the deduction of the B-integral contribution for a specific harmonic order, are presented in Table 1. The color bars represent the magnitudes of the harmonic **(A–B)** and fundamental **(C)** intensity signals, respectively.

Fig. 2A,B shows the 2D colour maps of the intensity- induced relative fringe shifts associated with a change in the dipole phase, $\Delta\varphi_{q,exp}$, for the H7 (A) and H9 (B) in the bulk MgO recorded as a function of the varying relative peak intensities of the NIR pulse replicas. In the intensity-dependent measurements, the interaction of intense 800nm laser pulses with the MgO solid resulted in an additional nonlinear response from the material. Since the refractive index of the medium varies with intensity due to the optical Kerr effect (*47*), it leads to a nonlinear modulation of the refractive index in MgO. This modulation induces a nonlinear phase shift in the light wavefront, known as the B-integral. Thus, the B-integral is a critical parameter in this context. In our interferometric arrangement, we took advantage of the setup's sensitivity to phase changes to measure the intensity-dependent non- linear phase shift in the solid sample (fig. S8). Fig. 2C

shows the 2D color map of the relative nonlinear phase shift, $\Delta\varphi_{800}$, accumulated by the 800 nm driver in the MgO within the same measured intensity range, as displayed in panels (A) and (B). By analyzing the corresponding interference patterns, we could gain insights into the nonlinear phase shift of the investigated medium and therefore deduce the B-integral contribution from the XUV measurements.

In Fig. 2, the $\Delta\varphi_{q,exp}$ and $\Delta\varphi_{800}$ values were determined from linear fits to the fringe shifts and the associated y-axis intercept over the 3–9 TW cm−2 change in peak intensities range. For the XUV data, the extracted $\Delta\varphi_{q,exp}$ values for the H7 and H9 are $0.96 \pm 0.09\ (2\pi) \cdot rad$ and $1.27 \pm 0.16\ (2\pi) \cdot rad$, respectively. For the 800 nm data, the $\Delta\varphi_{800}$ value of the accumulated B-integral in the MgO sample is determined to be $0.085 \pm 0.01\ (2\pi) \cdot rad$. Taking into account the harmonic order $q = 7$ and 9, the total B-integral contribution to the XUV phase, $\Delta\phi_{q,B}$, can be defined as: $\Delta\varphi_{q,B} = \Delta\varphi_{800} \cdot q$. Thus, the cumulative nonlinear phase shift in the 100 μm-thick MgO crystalline solid is found to be $\Delta\varphi_{7,B} = 0.6 \pm 0.01\ (2\pi) \cdot rad$, and $\Delta\varphi_{9,B} = 0.76 \pm 0.01\ (2\pi) \cdot rad$. By subtracting the $\Delta\varphi_{7,B}$ and $\Delta\varphi_{9,B}$ contributions of the B-integral from the $\Delta\varphi_{7,exp}$ and $\Delta\varphi_{9,exp}$ intensity-induced harmonic phase fringe shifts, we can derive the experimental relative dipole phase in radians as: $\Delta\varphi_q = \Delta\varphi_{q,exp} - \Delta\varphi_{q,B}$. As a result, the obtained $\Delta\varphi_7$ and $\Delta\varphi_9$ values are $2.28 \pm 0.56\ rad$ and $3.22 \pm 0.92\ rad$, correspondingly. In line with the previously reported studies by Lu *et al.* (31), the relative dipole phase in MgO increases linearly with intensity when examining XUV emission from H7 to H9 over the same range. Nevertheless, our findings go beyond those of previous results (31), showing that the nonlinear phase accumulated by the intense 800 nm pulses in the sample is markedly substantial – nearly two-thirds of the intensity-induced XUV phase shift recorded. Consequently, for a correct interpretation of intensity-dependent dipole phase data, it is vital to measure and subtract the B-integral contribution in every XUV interferometric experiment where nonlinear effects are significant.

## Discussion

**Theoretical simulations of the harmonic dipole phase.** To elucidate the fundamental physical mechanisms governing the dipole phase and aid in analyzing experimental data in MgO, we performed analytical and numerical calculations for two-band and multi-band systems (see Materials and Methods). We began with the implementation of an analytical two-band model that examines the semi-classical laser-driven $k$-dependent electron trajectories of the electrons in the conduction band (CB) $\epsilon_e^k$ and the holes in the valence band (VB) $\epsilon_h^k$, akin to those involved in gas-phase HHG. The semi-classical action with electron trajectories that have ionization times $t_i$ and recollision times $t_f$ was derived from the semiconductor Bloch equations (SBE) under the low electron inversion limit, utilizing the interband saddle-point approximation (*3*). The dipole phase of the emitted XUV light was then analyzed through the semi-classical action (*18,35*).

$$S(t_f) = \int_{t_i}^{t_f} \Delta\epsilon(k(\tau)) d\tau \quad \text{with} \quad \Delta\epsilon_k = \epsilon_e^k - \epsilon_h^k \tag{1}$$

Accordingly, from the semi-classical action, the dipole phase $\varphi_q$ of harmonic order $q$ can be obtained as (*35*):

$$\varphi_q = q\left(\omega_0 t_f + \frac{\pi}{2}\right) - S(t_f) \tag{2}$$

While this expression gives the phase contribution to HHG that has its origin in the interband current, which is typically dominant for above-band-gap emission, we show below that the intraband HHG contribution in many cases has a comparable phase profile. Furthermore, the analytical approach makes explicit use of well-defined trajectories, which are often not clearly discernible in more advanced simulations of solid HHG when examining the time-frequency profiles (Gabor transforms) of the HHG emission. We examined previously (*48*) by comparing numerical SBE simulations to the analytical model, which, despite this shortcoming, the analytical model does predict the same phases as more advanced numerical simulations that do not show clearly discernible trajectories. The physical interpretation of Eqs.1,2 reveals the origin of the dipole phase in solid-state HHG and is illustrated in Fig. 3A for two intensity-dependent dipole phase trajectories, $\varphi(I_1)$ and $\varphi(I_2)$. In the reciprocal k-space, electron-hole pairs are generated due to excitation prompted by the driving NIR field intensities, $I_1$ and $I_2$, which are accelerated along separate trajectories (short and long) in the crystal band structure before ultimately recombining in the VB. The accumulated phases of the emitted XUV light for each harmonic order $q$ and different trajectories are experimentally measured as the relative dipole phase $\Delta\varphi_{q,exp} = \varphi(I_1) - \varphi(I_2)$, which can also be computed. Fig. 3B presents simulation results derived from the two-band analytical model, depicting the dipole phase trajectories as a function of time for H7 and focusing solely on the short trajectories, with $\varphi(I_1)$ (yellow line) at an intensity of 6 TW cm$^{-2}$ and $\varphi(I_2)$ (green line) at an intensity of 15 TW cm$^{-2}$. The corresponding dipole phases are calculated using Eqs.7-8, which are shown as shaded areas integrated over time in Fig. 3C. The dipole phase thus maps out the time-integrated motion of the coherently excited electron-hole wave packets, which accelerates through k-space.

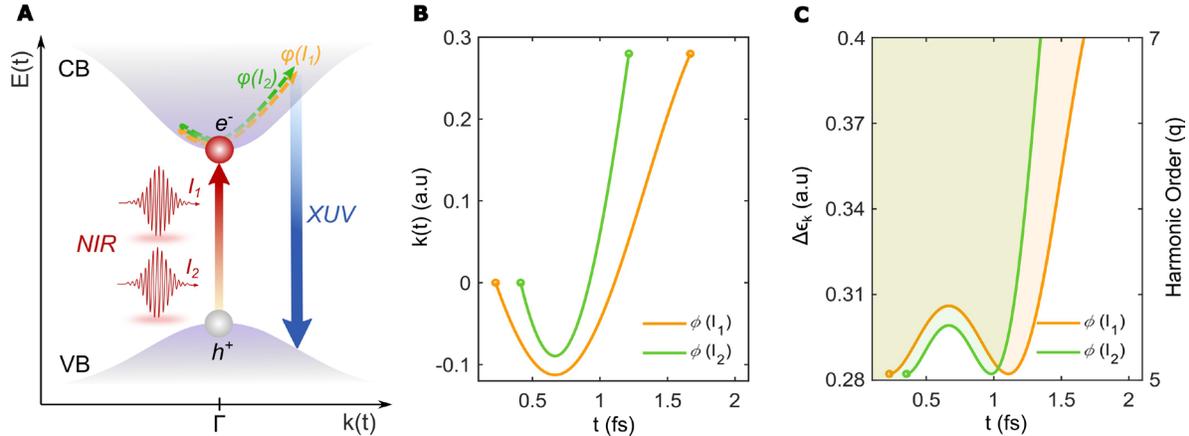

**Fig. 3. Dipole phase in solid-state HHG.** (**A**) Schematic illustration of the intensity-dependent dipole phase trajectories, $\varphi(I_1)$ and $\varphi(I_2)$, induced by the two NIR pulse replicas during HHG in MgO, depicted within the crystal band structure in reciprocal k-space. (**B**) Analytical semi-classical simulations of the dipole phase trajectories for H7 in MgO under applied driving NIR field intensities of 6 TW cm$^{-2}$ $\varphi(I_1)$ and 15 TW cm$^{-2}$ $\varphi(I_2)$. (**C**) Dipole phases $\varphi(I_1)$ and $\varphi(I_2)$ for H7, obtained using the analytical model shown as time-integrated shaded regions.

Next, the SBE were employed to numerically simulate the HHG process in MgO solid for a two-band system. The simulations involved using sparse spectral methods (*49*) on a one-dimensional Fourier basis to systematically calculate the microscopic (interband and intraband) contributions to the dipole phase. Comprehensive information regarding the numerical SBE-based simulations can be accessed in our previous publications (*48, 50, 51*). The use of Bloch states in two-band calculations, which exhibit complete spatial delocalization, motivates the exploration of an alternative basis comprising states that are tightly localized on specific lattice points, echoing the properties of atomic or molecular orbitals. Considering a realistic material electronic system involving multiple bands (or orbitals), we implemented a real-space, orbital-based framework and performed multi-band (full electronic band structure) numerical simulations of the dipole phase in MgO, using a basis derived from maximally localized Wannier orbitals (*52, 53*). Note that all theoretical simulations presented in this study were carried out for MgO solid along the Γ–X high symmetry crystal direction, which pertains to the Mg–O bond. This is the orientation in which experiments were performed. Additional simulations along the Γ–K–X direction (or Mg–Mg bond) have also been conducted and can be found in the Supplementary Materials.

Fig. 4 shows the experimental and calculated relative dipole phase as a function of the driving laser peak intensity, for H7 and H9 of the fundamental 800 nm field in MgO. For both $\Delta\varphi_7$ and $\Delta\varphi_9$, the analytical semi-classical calculations are presented for the short (s) and long (l) electron trajectories, while the numerical simulations based on the SBE two-band and the Wannier states multi-band models show no behavior indicative of exclusively short or long trajectories (*48*). Fig. 4A displays a comparison between the experimental dipole phase results for $\Delta\varphi_7$ (red dots) and $\Delta\varphi_9$ (blue dots) and three theoretical models: the analytical (s) model (pink dashed-dotted line), the numerical two-band model for interband (inter) transitions (orange tri- angles) and multi-band model (ruby solid line) for H7. For H9, the figure includes the numerical two-band model results for the intraband (intra) transitions (light blue stars) and the multi-band model (navy blue solid line). All comparisons are made within the measured intensity range from 3 to 9 TW cm$^{-2}$. The presence of an 800-nm-induced nonlinear phase (B-integral) in the MgO sample causes the linear fits of the experimental XUV fringe patterns within the inner and outer regions to diverge (see Fig. 2), resulting in relatively large error bars in Fig.4A, highlighted by the shaded red and blue areas for H7 and H9, respectively. Table 1 summarises the relative dipole phase values for H7 and H9 in MgO as obtained from experimental data and multi-band and two-band numerical models. The results clearly show that the theoretical dipole phase calculations for H7 in the low-intensity regime ($\leq 7.5$ TW cm$^{-2}$) are in good agreement with one another and with the experimental data, reflecting the largely semi-classical nature of short electron trajectories. At higher intensities, the multi-band effects likely become significant. Eventually, considering effects such as carrier-carrier and carrier-phonon scattering will become important, which are currently not included in our model. All named effects are especially important as the real-space trajectory length increases. This inherently suggests that the atomic dipole phase concept employed in gas-phase HHG is less applicable to solids under higher driving intensities. This is corroborated by the observation that the electronic band structure of MgO around the Γ-point remains approximately parabolic at low driving intensities (fig. S9), indicative of free-particle-like behavior in a low driving intensity setting. As the driving intensity increases, the effective momenta rise, enabling carriers to access energy levels that are offset from those of a free particle. In the case of H9, the experimental $\Delta\varphi_9$ behavior is almost flawlessly replicated by the numerical two-band and multi-band models. Conversely, the semi-classical approach fails to align with the experimental data due to the emission threshold of H7 being 5 TW cm$^{-2}$, compared to H9's higher requirement of around 15 TW cm$^{-2}$, as depicted in Fig. 4C. From a classical perspective, this can be attributed to the relationship between photon energy ($E_{ph} = E_g + E_{kin}$) and laser intensity, where $E_{kin} \propto k^2(t)$ increases with intensity.

In Fig. 4B and Fig. 4C, all computed results are shown for H7 and H9, respectively, extended to driving laser peak intensities up to 20 TW cm$^{-2}$. Fig. 4B displays additional findings for $\Delta\varphi_7$ simulated for intraband transitions via the two-band numerical model (yellow stars) and for long electron trajectories through the analytical model (purple dashed-dotted line). Furthermore, in Fig. 4C, we present the numerical results of the two-band model for interband transitions (blue-green triangles), alongside the semi-classical calculations for both short (light green dashed-dotted line) and long (dark green dashed-dotted line) trajectories for $\Delta\varphi_9$. The analytical model demonstrates that the dipole phase slope changes more rapidly for long trajectories than for short ones, a phenomenon attributed to the longer electronic quantum path present in gas-phase HHG, as reported in earlier studies (*11, 35*). Notably, the analytical model of the short trajectories agrees well with both experiment and other theories, while the long-trajectory analytical model is off. As explained above and in (*48*), solid-state HHG usually does not show any signatures of long-trajectory contributions, which is thus confirmed by the phase measurements. We also note that the interband and intraband phase contributions are different for H7 at low intensities and remarkably similar in the intermediate experimentally accessible intensity range. This measurement thus confirms the interband current as the origin for H7, in accordance with simulated amplitude contributions (not shown. The similar phase accumulation for interband and intraband current likely has its origin in the few-band nature of HHG from MgO in the observed range. While interband harmonics are inherently sensitive to interband phases and thus to energy differences between bands, the intraband harmonics will be sensitive to intraband energy differences as a function of carrier momentum, which will result in a similar phase if the VB and CB are both contributing with a similar emission amplitude and band curvature. The dominant contribution of VB and CB to H7 and H9 in the measured intensity range is further corroborated by decomposing the phase contributions into their molecular orbital origins in Fig. 5.

**Table 1. Comparison of experimental ($\Delta\varphi_q$) and calculated ($\Delta\varphi_{q,calc}$) relative dipole phase values for harmonics 7 and 9 in MgO within the intensity range of 3 to 9 TW cm$^{-2}$, using numerical multi-band and two-band models.**

| Harmonic order (q) | $\Delta\varphi_q$ (rad) | $\Delta\varphi_{q,calc}^{multi-band}$ (rad) | $\Delta\varphi_{q,calc}^{two-band(inter)}$ (rad) | $\Delta\varphi_{q,calc}^{two-band(intra)}$ (rad) |
|---|---|---|---|---|
| 7 | 2.28 ± 0.56 | 1.48 | 1.02 | 0.37 |
| 9 | 3.22 ± 0.92 | 2.17 | 1.48 | 2.61 |

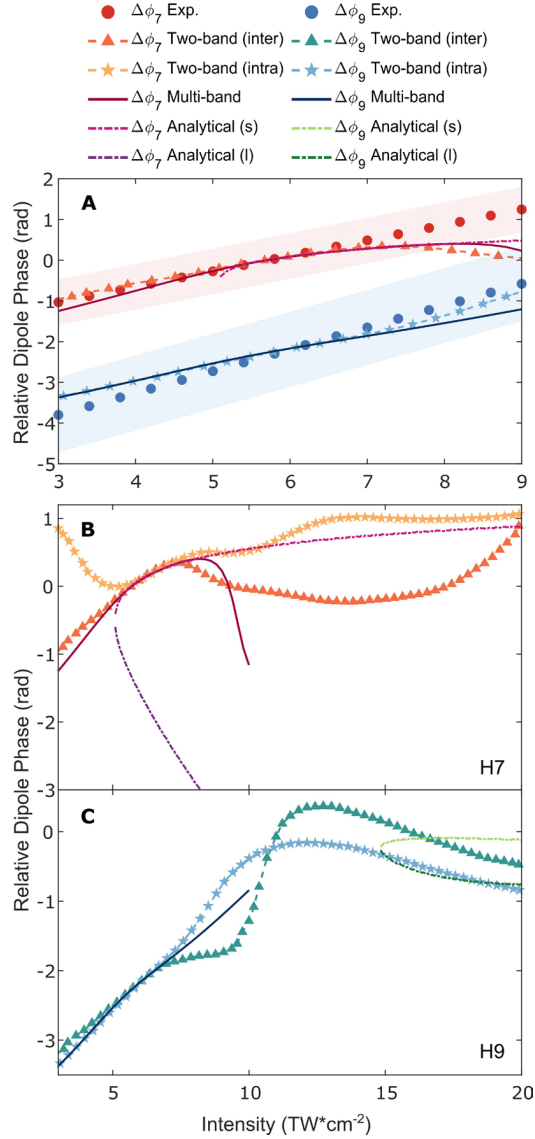

**Fig. 4. Intensity-induced relative dipole phase for harmonics 7 ($\Delta\varphi_7$) and 9 ($\Delta\varphi_9$) in MgO with the 800 nm driver. (A)** Experimental $\Delta\varphi_7$ (red dots) and $\Delta\varphi_9$ (blue dots) results for MgO along the Γ–X crystal direction (Mg–O bond) in comparison with the analytical semi-classical and numerical two-band, as well as multi-band simulations, within the measured intensity range of 3–9 TW cm$^{-2}$. The error bars, indicated by shaded areas, are derived from the linear fits of the experimental fringe patterns. **(B–C)** The simulated $\Delta\varphi_7$ (H7) and $\Delta\varphi_9$ (H9) dipole phase results, shown in panel **(A)**, extended to higher driving laser peak intensities ranging from 3 to 20 TW cm$^{-2}$, encompassing all computational findings. Note that the analytical calculations consider only 1 optical cycle (OC) of the 800 nm laser field, whereas the numerical two-band analysis accounts for 8 OC. For optimal comparison of the theoretical models, the curves are also vertically offset, with a reference point set at 6 TW cm$^{-2}$ for both H7 and H9.

**Orbital-resolved contributions to the harmonic dipole phase.** We delved deeper into the physical origins underlying the dipole phase by assessing the various orbital contributions to the harmonic dipole phase in MgO, using an orbital-based framework on the Wannier basis. Fig. 5 illustrates the contributions of three magnesium (Mg) orbitals (s, p, d) and one oxygen (O) orbital (p) to the relative dipole phase of harmonics H7 (A) and H9 (B), as determined by Fourier phases and amplitudes calculated over an intensity range of 3–10 TW cm$^{-2}$ through multi-band numerical analysis. The results reveal several key findings: For H7, the dipole phase at lower driving intensities (<8 TW cm$^{-2}$) is mainly affected by nearly equal contributions from the Mg (s) and O (p) orbitals. However, at higher applied intensities, the respective contributions of the Mg (s) and O (p) orbitals decrease, and the contributions of the Mg (p) and Mg (d) orbitals become prevalent. For H9, the phases of all orbitals increase linearly with intensity, with the O (p) orbital making the largest contribution, thereby primarily determining the overall harmonic dipole phase. The dominant contributions of Mg (s) and O (p) orbitals in MgO are not unexpected, as they have been identified as the primary contributors to the material's electronic properties in the literature (*54*). As evident by the projected density of states (PDOS) simulations (fig. S9), the highest occupied molecular orbital (HOMO) is dominated by the O (p) orbitals, whereas the lowest unoccupied molecular orbital (LUMO) is dominated by Mg (s). This coincides with the intuitive ionic bonding view where the more electronegative oxygen atom attracts two electrons from magnesium. Therefore, the measured phases at low and moderate intensities of both H7 and H9 are dominated by electron dynamics in HOMO and LUMO of MgO. At higher intensities the contributions from orbitals other than HOMO and LUMO become dominant as the electrons can be promoted into those orbitals too. This orbital picture on the origin of harmonic phases is equivalent to the more common band-structure view of solid-state HHG, where excitation prepares coherent electron-hole pairs, and laser-driven electron dynamics further promote these excited carriers into higher and lower lying bands. Complementary simulations for harmonic 5 (H5, 8 eV) in MgO, not observed experimentally in this study but computed (see fig. S11), further reinforce the notion that Mg (s) and O (p) orbitals significantly contribute to the dipole phase, including within the band-gap regime.

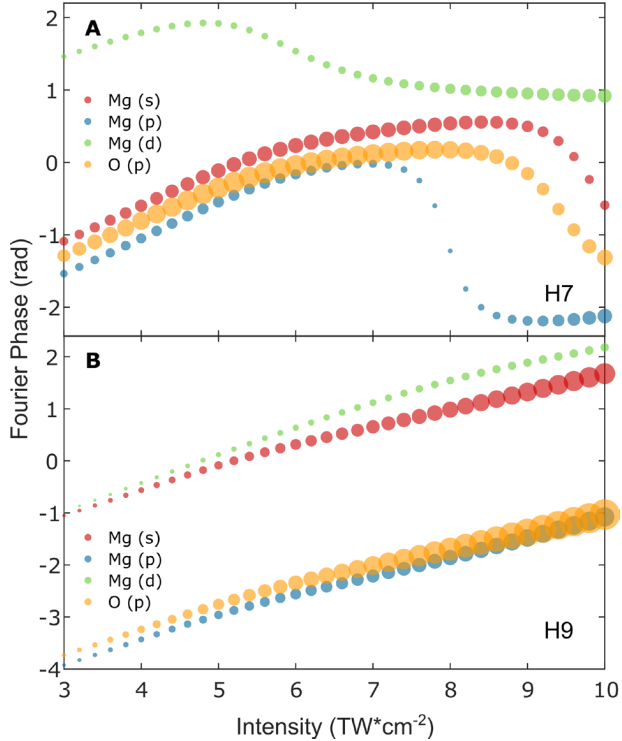

**Fig. 5. Orbital contributions to the harmonic relative dipole phase in MgO.** The Fourier phases, weighted by the amplitudes of the three magnesium Mg (s, p, d) and one oxygen O (p) orbitals as a function of the driving laser peak intensity for H7 (**A**) and H9 (**B**) in MgO, based on the numerical multi-band simulations. The size of each circle corresponds to the Fourier amplitude associated with the individual orbital, while color coding is used to distinguish between them visually. The results are displayed for the 800 nm driving laser pulses aligned along the Mg–O bond in the Γ–X crystal direction of MgO.

**Conclusions and outlook.** This study has provided a detailed examination of the intensity-dependent dipole phase in solid-state HHG through the use of ultrastable XUV interferometry. The synthesis of experimental results with multifaceted theoretical insights into the harmonic phase in MgO has yielded the following key findings: the generation of high-harmonics using the 800 nm driver in the above-band-gap regime suggests that the interband mechanism plays a key role in the emitted phase. Moreover, the insights gained from this work not only deepen our understanding of the fundamental processes involved in HHG from solids but also pave the way for better control over macroscopic phase matching and the divergence and focusing characteristics of the emitted XUV light. This is particularly enabled by establishing and experimentally verifying an analytical model for the dipole phase, which can serve as input for simulations of phase matching and HHG wavefronts.

By carefully manipulating the driving laser intensity, one can achieve attosecond-level tuning of the harmonic phase in solid-state HHG. This could have profound applications in fields such as attosecond science, quantum optics, and high-resolution imaging. In addition, our approach not only facilitates the direct measurement of the harmonic dipole phase but also encapsulates sensitivity to the B-integral, thereby providing a robust framework for exploring nonlinear effects in real-time. Future research should aim to explore how these nonlinear effects influence HHG in a range of materials with distinct optical properties, including nonlinear crystals, amorphous solids, and nanostructures, as this could facilitate the development of compact, efficient, and stable all-solid-state XUV light sources, providing high-brightness light for probing intricate material properties at the nanoscale.

## Materials and Methods

### Experimental details

A Ti: Sapphire laser amplifier system (Astrella, Coherent) was used to produce NIR pulses with an 800 nm central wavelength (1.55 eV photon energy), 7 mJ pulse energies, 45 fs pulse durations, at a 1 kHz repetition rate. To generate the ultrashort NIR pulse replicas, 4 mJ pulse energies were driven to a birefringent common-path interferometer assembled from two pairs of alpha-barium borate wedges. The resulting two pulse replicas were directed to a f = 500 mm spherical mirror, which focused the beams into a solid target to generate the XUV pulses through the HHG process. The double-side polished, 100 $\mu$m-thick, (100)-cut bulk crystalline magnesium oxide sample (purchased from Crystal GmbH) was chosen as a solid HHG target. In the experiments, the linearly polarized driving laser pulses were aligned with their polarization direction fixed along the Γ—X direction (Mg–O bond) of the solid. Focusing the NIR pulse replicas, derived from attenuated 11 μJ pulse driving laser energies into the sample, led to a spatio-temporal peak intensity of 6 TW cm$^{-2}$, when the intensities at each focus were equal. The two foci of NIR pulse replicas with a spot size of 50 μm (FWHM) were spatially separated vertically on the MgO target by the distance being approximately twice the beam spot diameter (190 μm) (fig. S2). After generating two XUV pulses from both foci, they were spatially overlapped in the far-field in a vacuum detection chamber. An aberration-corrected, concave, flat-field, diffraction grating (Shimadzu, 1200 grooves/mm), housed in the detection chamber, was used to spectrally disperse the harmonics onto a double-stack multi-channel plate (MCP) and a phosphor screen detector (Photonis USA). The back of the phosphor screen was imaged by a CMOS camera (Basler ace, acA1300-200μm), mounted outside the detection chamber. For HHG and XUV interferometry, we employed a low-vibration turbomolecular pump (Edwards, STP-XA3203C), which was backed by a scroll pump (Pfeiffer, SEK 28/40), maintaining high pressures in the vacuum chambers below 10$^{-8}$ mbar during the experiments. A schematic of the experiment can be found in the Supplementary Materials (fig. S1).

**Numerical and analytical models**

We simulate the dipole phase in MgO along a single high-symmetry crystal direction: Γ–X (Mg–O bond) or Γ–K–X (Mg–Mg bond), utilizing either two-band or multi-band (full band structure) models in conjunction with semiconductor Bloch equations (SBE). In the numerical and analytical two-band models, we employ the electronic band structure of cubic MgO (100) for a two-level system that includes the first valence band (VB) and the first conduction band (CB). The computations are performed for the MgO crystal structure, which has a band gap of $E_g$ = 7.8 eV (55), a lattice constant of $a$ = 4.19 Å, a VB height of 1.1 eV, a CB height of 5.5 eV, a transition dipole moment at the Γ-point of 0.78 a.u. and a dephasing time of 3 fs. In the numerical two-band model, one high-symmetry crystal direction of MgO (either Γ–X or Γ–K–X) is represented by a one-dimensional chain, with the assumption of tight-binding sinusoidal bands. The system is driven well below resonance using 800 nm ($\tau_{800} = 2.66$ fs) Gaussian envelope laser pulses consisting of 8 optical cycles (OC). The specifics of our model are detailed in prior publications (*48, 50, 51*). The analytical semi-classical model is derived from the laser-driven electron trajectories. This is achieved using the SBE by applying the saddle-point approximation to the interband current while assuming low carrier inversion (*3*). The trajectories are computed for 1 OC of the driving 800 nm laser field. Note that the intraband current is completely neglected in this model.

To obtain the trajectories, we consider the time-dependent effective carrier momentum during acceleration, which follows that of the driving field as:

$$k(t) = A(t) - A(t_i) + k_0 \tag{3}$$

where $t_i$ is the excitation time and $A(t)$ is the vector potential. We consider a cosine driving field which has a corresponding vector potential of

$$A(t) = \int E(t)dt = -\frac{E_0}{\omega_0}\sin(\omega_0 t) \tag{4}$$

Here, we neglect tunneling and restrict excitation to the Γ-point ($k_0 = 0$), resulting in distinct long and short trajectories, similar to those observed in gases (*56*). Note that these distinct trajectories are not present in the numerical SBE simulations, as they facilitate excitation across the entire k-space of the bands.

The real-space trajectories per band $\lambda \in \{e, h\}$ for the different $t_i$ are obtained by integration of the carrier group velocity $v_\lambda = \delta_k \epsilon_\lambda^k$ over time, resulting in

$$x_\lambda(t) = \int_{t_i}^{t} v_\lambda(k(\tau))d\tau \tag{5}$$

The group velocity is derived from the band structure of MgO, adopting the same two-band configuration as in the numerical SBE simulations.

As the real-space distance between the electron and hole reaches zero, recombination with the time $t_f$ occurs, establishing the condition:

$$\Delta x(t_f) = x_e(t_f) - x_h(t_f) = 0 \tag{6}$$

Only the trajectories that undergo recombination contribute to the interband current, with the corresponding photon energy being the energy difference between the charge carriers at the time of their recombination. Thus, the analytical semi-classical model is limited to evaluating harmonics that have energies greater than the bandgap but less than the maximum energy difference between the CB and VB.

As discussed earlier in the main text, the phase of the emitted XUV light for each trajectory can be represented by the semi-classical action (*18*):

$$S(t_f) = \int_{t_i}^{t_f} \Delta\epsilon(k(\tau)) d\tau \quad \text{with} \quad \Delta\epsilon_k = \epsilon_e^k - \epsilon_h^k \tag{7}$$

Therefore, the dipole phase for harmonic order $q$ can be derived from the semi-classical action as follows (*35*):

$$\varphi_q = q\left(\omega_0 t_f + \frac{\pi}{2}\right) - S(t_f) \tag{8}$$

In the numerical multi-band model, we first calculate the field-free band structure and dipole couplings for MgO through density functional theory (DFT) using the electronic band structure code Quantum Espresso (*57*). We use the Heyd–Scuseria–Ernzerhof exchange-correlation hybrid functional on a Monkhorst–Pack (MP) grid of 10×10×10 points. The Bloch states are projected onto a set of maximally localized Wannier functions utilizing the s, p, and d orbitals of magnesium (Mg) and the p orbitals of oxygen (O) with the code Wannier90 (*58*). The conduction bands are shifted to match the experimental band gap of $E_g$ = 7.8 eV (*59*). The calculated band dispersion and the orbital-resolved projected density of states are presented in fig. S9. The time-dependent propagation is subsequently performed employing the density matrix formal- ism with the code detailed in (*52*), incorporating the experimental laser parameters (intensity, frequency, and pulse duration). Decoherence effects are included phenomenologically via a dephasing time $T_2$ = 3 fs. We observe only small shifts to the dipole phase when changing $T_2$ from 3 fs to 6 fs in the harmonic and intensity range of interest. The orbital contributions to the current are extracted following the procedure in (*53*).

# Acknowledgments

**Funding:** Part of this work was conducted at the Advanced Research Center for Nanolithography (ARCNL), a public private partnership between the University of Amsterdam (UvA), Vrije Universiteit Amsterdam (VU), Rijksuniversiteit Groningen (RUG), the Dutch Research Council (NWO), and the semiconductor equipment manufacturer ASML, and was partly financed by 'Toeslag voor Topconsortia voor Kennis en Innovatie (TKI)' from the Dutch Ministry of Economic Affairs and Climate Policy. This manuscript is part of a project that has received funding from the European Research Council (ERC) under the European Union's Horizon Europe research and innovation programme (Grant Agreement No. 101041819, ERC Starting Grant ANACONDA). The manuscript is also part of the VIDI research programme HIMALAYA with Project No. VI.Vidi223.133 by NWO. R.E.F.S. acknowledges support from Fellowship No. LCF/BQ/PR21/11840008 from "La Caixa" Foundation (ID 100010434). This research was supported by Grant No. PID2021-122769NB-I00 funded by MCIN/AEI/10.13039/501100011033. A.J.G. acknowledges support from the Talento Comunidad de Madrid Fellowship 2022-T1/IND-24102 and the Spanish Ministry of Science, Innovation and Universities through grant reference PID2023-146676NA-I00.

**Author contributions:** N.K. and P.M.K. conceived the research. N.K. designed, performed the experiments, and analyzed the data. P.J.v.E. and B.d.K. developed the two-band theory and carried out the analytical and numerical simulations under the supervision of P.M.K. A.J.G. and R.E.F.S. developed the multi-band theory and performed the numerical calculations. N.K. and P.M.K. wrote the manuscript. All authors contributed to discussions and the interpretation of the results.

**Competing interest:** The authors declare that they have no competing interests.

**Data availability:** All data needed to evaluate the conclusions in the paper are present in the paper and/or the Supplementary Materials.

# Supplementary Materials for

## Attosecond high-harmonic interferometry probes orbital- and band-dependent dipole phase in magnesium oxide

Nataliia Kuzkova *et al.*

Corresponding authors: n.kuzkova@arcnl.nl (N.K.); p.kraus@arcnl.nl (P.M.K.)

**The PDF file includes:**

    Supplementary Text
    Figs. S1 to S13
    References

# Supplementary Text

**XUV interferometric setup**

Fig. S1 illustrates a schematic overview of the main components of the XUV interferometric setup. The Materials and Methods section of the main text provides a comprehensive description of the laser system and its key components. Here, we focus on the details of a birefringent common-path interferometer, which generated the ultrashort near-infrared (NIR) pulse replicas. The propagation of the fundamental NIR beam through and after the interferometer, including polarization direction diagrams (red arrows), is illustrated in the inset of Fig. S1. The interferometer consists of two pairs of 20x20x1 mm alpha-barium borate ($\alpha$-BBO) birefringent wedges (United Crystals Inc.), with an optical axis 45° away from 20 mm edges and an apex angle, $\alpha = 14°$, which corresponds to a relatively large birefringence, $\Delta n = 0.1189$ at 800 nm wavelength, where $\Delta n$ is the difference between the refractive indices of extraordinary, $n_e$, and ordinary, $n_o$, polarization, respectively. Orientation of the optical axis of the first pair of wedges (W1-2) at an angle of 45° with respect to the input linearly polarized NIR beam (vertical polarization with respect to the laser table) resulted in the generation of two orthogonally polarized phase-locked collinear replicas along the propagation direction. These two pulse replicas with the same intensity distribution were then sent through a second pair of wedges (W3-4) whose optical axis was perpendicular to the W1-2. As a result, the relative delay, $\Delta\tau$, between the two orthogonally polarized pulse replicas was introduced by the interferometer, which can be expressed as [1]:

$$\Delta\tau = \frac{\Delta n \cdot \Delta x}{c} tan\alpha$$

where $\Delta x$ represents the displacement of one of the wedges achieved by shifting it transversely relative to the laser beam and changing its thickness, and $c$ is the speed of light in vacuum.

By displacing the last wedge (W4), which was mounted on a linear piezo stage (SmarAct, SLC-1740s, 26 mm travel range, 1 nm resolution), the sub-20-attosecond $\Delta\tau$ interferometric temporal precision was achieved. The two orthogonally polarized NIR pulse pairs produced after the interferometer were subsequently aligned to the same linear polarization axis by utilizing a thin film polarizer (TFP). A half-wave plate (HWP) placed in front of the interferometer was used to control the orientation of the input polarization axis.

The size and distance between the two spatially separated NIR beams on the MgO sample were defined based on separate measurements performed in the air using a beam profiling camera (Gentec-EO, BEAMAGE-4M). To achieve this, a pickoff silver mirror (PM) was positioned before the vacuum HHG chamber to reflect the two attenuated fundamental laser beams onto the camera (see Fig. S1). The focal spots of the NIR beams were adjusted by slightly tilting the first wedge (W1), thus controlling the separation distance between them. Fig. S2 shows the spatial intensity profiles of the NIR focal spots measured as a function of the transverse position, resulting in each NIR beam focal spot size of a 50 μm (full-width at half-maximum, FWHM). The two-foci separation distance was determined to be 190 μm, which is nearly twice the beam spot diameter. In our time-dependent studies of the harmonic relative phase, we observed that the influence of the fundamental 800 nm frequency component on the harmonic phase becomes substantial when the distance between the NIR focal spots is < 1.5 times the beam spot diameter. To prevent fundamental frequency oscillations and guarantee that harmonic signals originate from independent foci, we consistently maintained a distance of twice the beam spot diameter between the NIR foci in all our experiments.

**Time-dependent harmonic relative phase shift measurements**

The temporal far-field XUV interferograms obtained from MgO, measured at equivalent peak intensities at each NIR focus of 6 TWcm$^{-2}$, are presented in Fig. S3. The recorded interferograms provided insight into the time-dependent relative phase shifts of H7 (refer to the main text) and H9, allowing us to analyze the sensitivity of the birefringent common-path interferometer to the harmonic phase changes.

Fig. S4 shows the 2D colour map representing the time-dependent relative phase shift of H9 in MgO, recorded as a function of the time delay, $\Delta\tau$, between the NIR pulse replicas. In the solid-phase HHG experiments, the NIR beams peak intensities of several TWcm$^{-2}$ were applied, constrained by the material damage threshold, which was ≈11 TWcm$^{-2}$ in this study. Consequently, the emission signals for higher harmonic orders, such as H9 in this instance, exhibit a lower signal-to-noise ratio compared to H7, as displayed in Fig. S3. Nonetheless, we successfully extracted the harmonic phase shift value for H9 within a 2.7 fs range (800 nm laser field optical cycle), which was found to be $\Delta\varphi_9^\tau = 9$ ($2\pi$*rad). This value corresponds exactly to the harmonic order q = 9, confirming the precise synchronization of high-harmonics with the NIR driving wavelength.

**Intensity-induced harmonic relative dipole phase measurements**

Fig. S5(A) shows the harmonic yield in MgO for H7 and H9 as a function of varying relative NIR foci peak intensities, scanned from 3 to 9 TWcm$^{-2}$. Panels (B-C) show the intensity-dependent profiles for the photon-energy integrated H7 and H9 interference fringes within MgO at 3, 6, and 9 TWcm$^{-2}$, highlighting the impact of varying intensities on the interference pattern. These data were used in evaluating intensity-induced relative dipole phase shifts, $\Delta\varphi_{q,exp}$, presented in Fig.2(A-B) of the main text. First, each fringe line-out of the intensity-dependent interference pattern profile for H7 and H9 was analyzed using a nonlinear least-squares algorithm to fit a Gaussian envelope. The identified maxima of the individual line-outs, corresponding to the peak spectral intensity values for each harmonic interferogram, were utilized to transform the divergence x-axis measured in milliradians (mrad) into the fringe shift axis in radians (rad). In this context, the locations of maxima and minima within the fringe pattern signify a ($2\pi$*rad) phase shift, reflecting the distance between two line-out peaks and their relative displacement.

Fig. S6 exemplary displays the raw data measured at 6 TWcm$^{-2}$, fitted to Gaussian envelopes for H7 (left panel) and H9 (right panel) in MgO, with the x-axis converted to represent the relative fringe shift in ($2\pi$*rad). From the fitted data, the extracted fringe maxima line-out points for H7 and H9 (indicated by the red and blue dots, respectively, in Fig. 2(A-B) of the main text) were then used to perform a linear fit of the intensity-induced relative fringe shifts within the 3 – 9 TWcm$^{-2}$.

**Intensity-induced nonlinear phase shift (B-integral) measurements**

To isolate and exclude any additional nonlinear effects in MgO that might occur due to the interaction of intense 800 nm laser pulses with the material, we conducted separate intensity-induced phase shift measurements that only recorded the response of the MgO nonlinear medium to the fundamental beam. In the 800 nm interferometric experiments, we ensured that the conditions were identical to those employed in the XUV experiments, up to focusing within the MgO solid. After the NIR beams propagated through the MgO sample, the images of horizontal interference fringes at 800 nm were directly recorded on the CMOS camera, set up in a transmission geometry in the far-field, following the removal of the XUV grating from the beam path (see Fig. S1). The far-field interferograms were recorded as a function of the HWP rotational position, corresponding to variations in the peak intensities of the NIR foci over the same intensity range used in the XUV experiments (from 3 to 9 TWcm$^{-2}$). Fig. S7 shows the far-field interferogram of the fundamental 800 nm driving wavelength obtained from MgO, presented when the peak intensities of the NIR beams were matched at 6 TWcm$^{-2}$ for each focus.

The left panel in Fig. S8 displays the intensity-dependent profiles of interference fringe patterns at 800 nm obtained from MgO, presented for the NIR foci peak intensities of 3, 6, and 9 TWcm$^{-2}$. While the relative fringe shift for varying fundamental intensities is not as prominent as observed in the XUV data (see Fig. S5(B-C)), it gains importance when we recognize that this phase shift scales linearly with the harmonic order. Analogously to the XUV data analysis, each fringe line-out of the intensity-dependent interference pattern profile at 800 nm was fitted to the Gaussian envelope, as exemplary shown in the right panel of Fig. S8. Using the fitted data, the extracted fringe maxima line-out points (shown as gray dots in Fig. 2(C) of the main text) were employed to carry out a linear fit of the intensity-induced nonlinear phase shift, $\Delta\varphi_{800}$, across the identical measured intensity range of 3 to 9 TWcm$^{-2}$ as in the harmonic data. This enabled us to quantify the $\Delta\varphi_{800}$ value in our XUV interferometric experiments and to subtract the B-integral contribution that had accumulated in the MgO sample at 800 nm from the XUV data.

**Multi-band (full electronic band structure) numerical calculations**

*1. Simulation results for the Γ–K–X crystal direction (Mg–Mg bond)*

The simulated intensity-induced relative dipole phase in MgO along the Mg–Mg bond (Γ–K–X crystal direction), based on the multi-band model, is depicted in Fig. S10, using a methodology analogous to that employed in the simulations presented for the Mg–O bond in the main text. Panel (A) displays the analysis results for harmonics 5 ($\Delta\varphi_5$), 7 ($\Delta\varphi_7$) and 9 ($\Delta\varphi_9$) of the fundamental 800 nm laser field for the intensities ranging from 3 to 10 TWcm$^{-2}$. Panels (B-D) illustrate the Mg (s, p, d) and O (p) orbital contributions to the relative dipole phase of H5, H7, and H9.

*2. Additional simulation results for the Γ–X crystal direction (Mg–O bond)*

Fig. S11 shows the additional results for the intensity-induced relative dipole phase, $\Delta\varphi_5$, simulated for H5 with the 800 nm driver along the Mg–O bond (Γ–X crystal direction) of MgO, using the multi-band model. A comparison of the intensity dependence of $\Delta\phi5$ for H5 versus H7 ($\Delta\varphi_7$) and H9 ($\Delta\varphi_9$) is shown in panel (A). The orbital contributions to the relative dipole phase of H5 are visualized in panel (B), using the calculated Fourier phases and amplitudes.

**Two-band analytical and numerical calculations**

For the two-band analytical and numerical calculations of the dipole phase the electronic band structure of MgO with a cubic (100) crystal structure was utilized, as described in Ref. [2]. The corresponding conduction (A) and valence (B) bands of MgO, as well as a cross-section of the bands along the two crystal directions of high symmetry: Γ–X (Mg–O bond) and Γ–K–X (Mg–Mg bond) are shown in Fig. S12. For both high symmetry directions, we assessed the dependence of the dipole phase on the fundamental 800 nm laser intensity. Panel (A) in Fig. S13 displays the analytically simulated intensity-dependent relative dipole phase results for harmonics 7 ($\Delta\varphi_7$) and 9 ($\Delta\varphi_9$), presented for both short (s) and long (l) electron trajectories, generated by the 800 nm laser pulses along the Γ–K–X crystal direction (Mg-Mg bond) in MgO. Panel (B) displays the corresponding results for the Γ–X direction (Mg-O bond), which are also addressed in the main text. According to the simulated analysis, the intensity-dependent dipole phase behavior of H7 exhibits similarities for both crystal directions involved in harmonic generation from MgO, with emissions initiating at about 5 TWcm$^{-2}$ peak intensity. For H9, harmonic emission begins at 11 TWcm$^{-2}$ for the Γ–K–X crystal direction and at 15 TWcm$^{-2}$ for the Γ–X direction. The relative dipole phase shift for the Γ–K–X direction is more pronounced compared to the other high-symmetry crystal direction in the same intensity range. Moreover, the larger relative dipole phase change is observed for the calculated long electron trajectories when compared to the short ones.

# Supplementary Figures

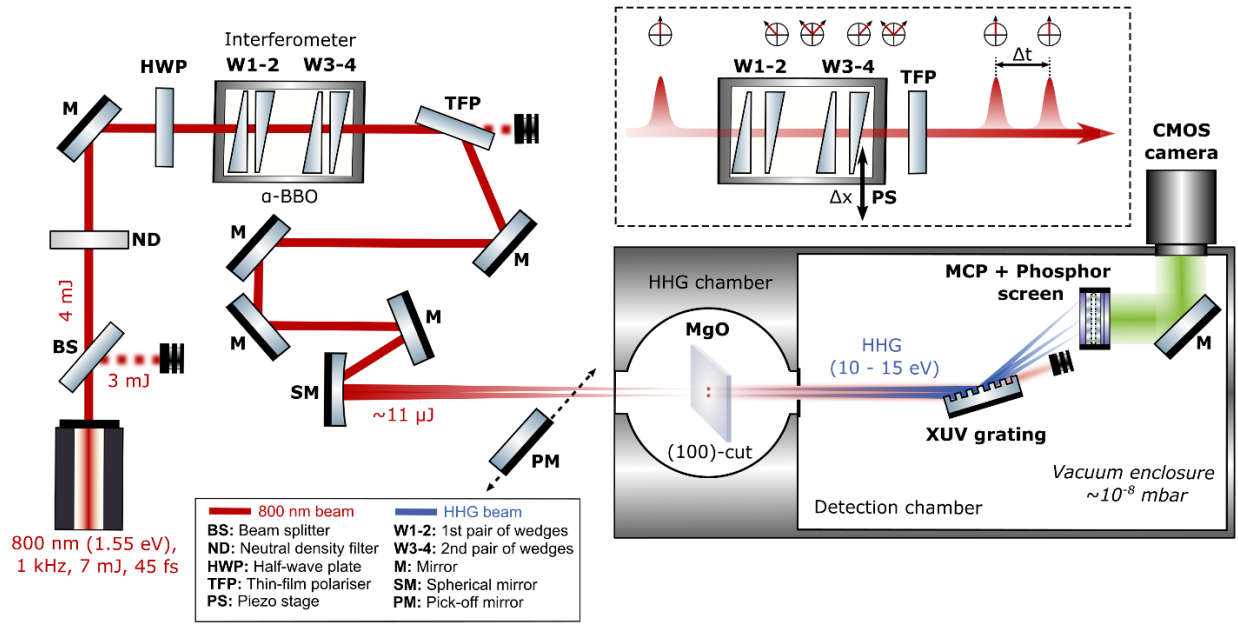

**Fig. S1. Schematic overview of the main components of the XUV interferometric setup.**

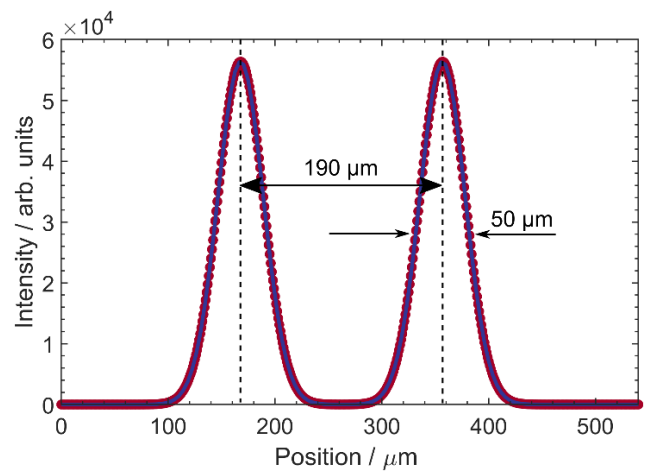

**Fig. S2. Spatial intensity profiles of the NIR foci.**
Spatial intensity profiles of the two NIR, used in the spectrally-resolved interferometric experiments on the solid, fitted by the Gaussian function (blue line), resulting in each beam focal spot size of 50 µm (FWHM). The peak positions (dashed lines) were used to reveal the vertical separation distance of 190 µm between the two beams, depicted by the black arrow in the figure.

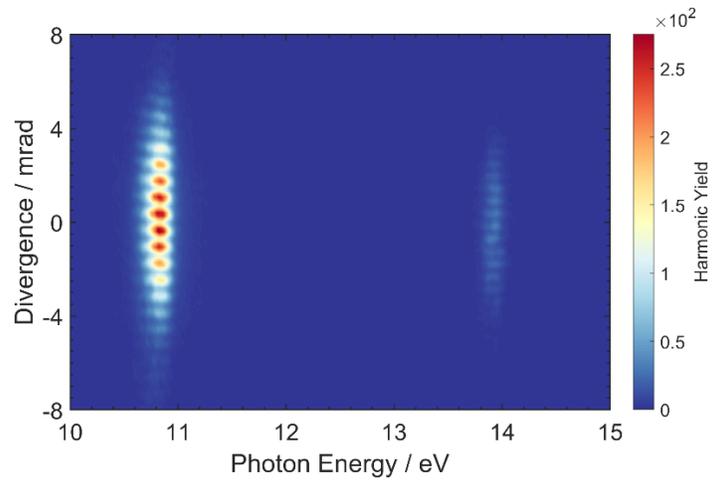

**Fig. S3. Temporal XUV interferograms obtained from MgO.**

The measured far-field interferograms for H7 (10.8 eV) and H9 (13.9 eV) at equivalent peak intensities at each NIR focus of 6 TWcm$^{-2}$. The harmonic orders (H7 and H9) correspond to the odd multiples of the fundamental driving frequency.

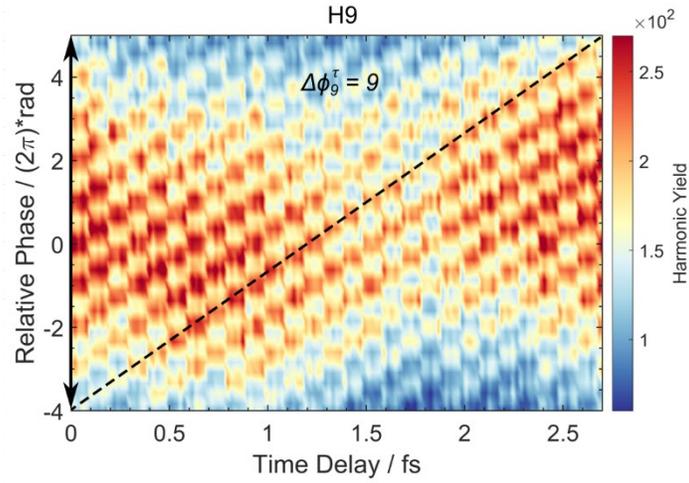

**Fig. S4. Time-dependent relative phase shift, $\Delta\varphi_q^\tau$, in MgO for harmonic 9 as obtained from the temporal XUV interferograms.**

The relative fringe shift of H9 within one optical cycle (2.7 fs) of the 800 nm driving laser field is shown with a dashed line. The extracted $\Delta\varphi_9^\tau$ value within this range is shown by the black arrow and is also depicted in the figure.

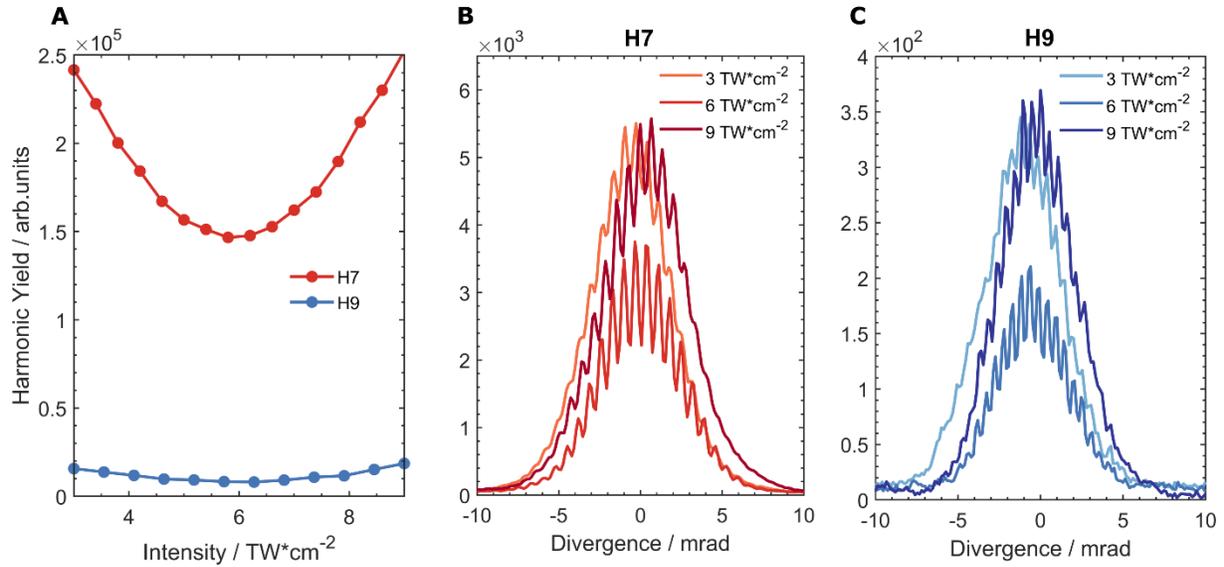

**Fig. S5. Intensity-dependent profiles of interference fringes for harmonics 7 and 9 in MgO.** The harmonic yield in MgO for H7 (red) and H9 (blue) as a function of NIR foci peak intensities, as recorded for 3–9 TWcm$^{-2}$ range in the intensity-induced dipole phase measurements (A). The profiles of photon-energy-integrated H7 and H9 interference fringe patterns at 3, 6, and 9 TWcm$^{-2}$, highlighting the dependence of phase shift on laser intensity (B–C).

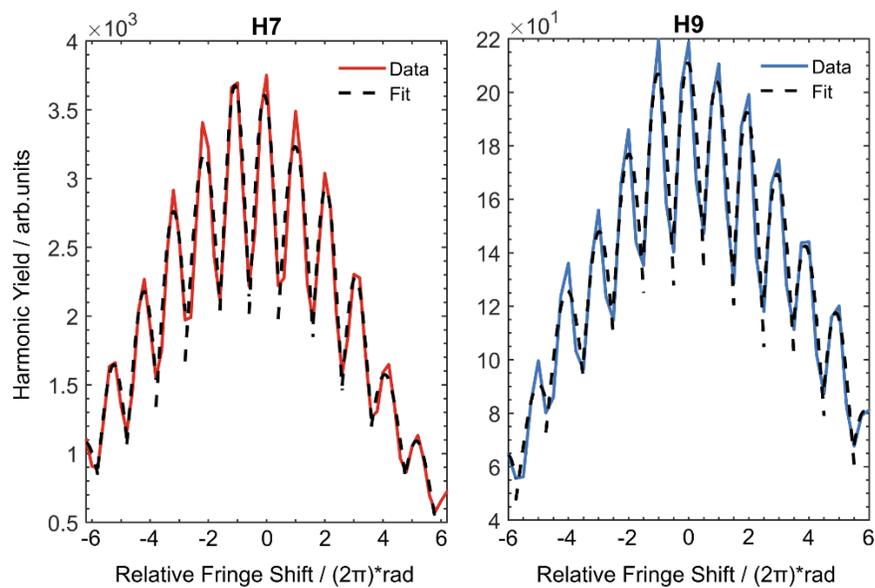

**Fig. S6 XUV data fitting procedure.**

The raw data line-out fringes from the intensity-dependent XUV interferograms (solid lines) presented for the NIR foci peak intensities of 6 TWcm−2, fitted to Gaussian envelopes (dashed lines) for H7 (left panel) and H9 (right panel) in MgO. The distance between each fringe maxima line-out corresponds to a (2π*rad) phase shift.

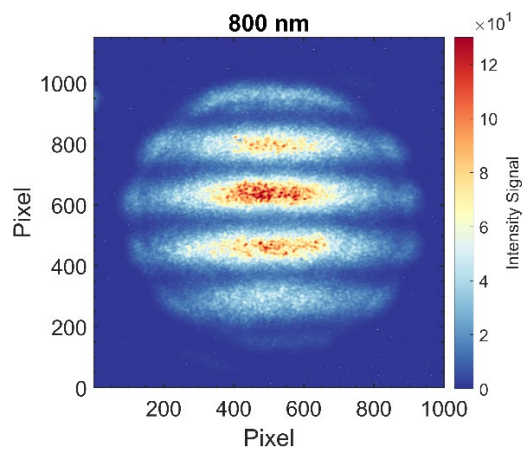

**Fig. S7 Intensity-induced interferogram for 800 nm emission from MgO.**

The far-field interferogram for the fundamental 800 nm driving field obtained from MgO during the intensity-dependent (B-integral) measurements, presented for the intensities of the NIR foci set to be the same (6 TWcm$^{-2}$).

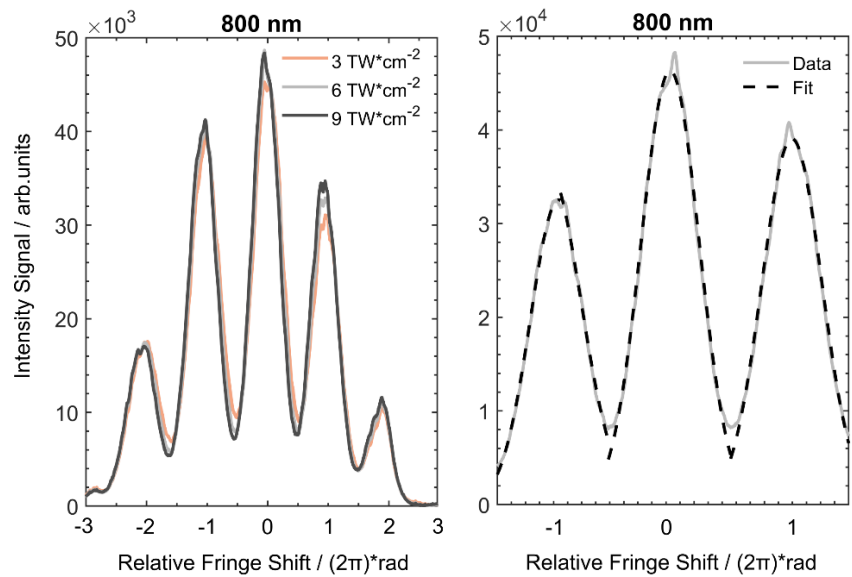

**Fig. S8 800 nm (B-integral) data fitting procedure.**

The intensity-dependent profiles of interference fringe patterns at 800 nm obtained from MgO, presented for the NIR foci peak intensities of 3, 6, and 9 TWcm$^{-2}$ (left panel). The selected line-out data of fringes (solid line) from the nonlinear phase shift measurements (refer to Fig. 2C in the main text), displayed for 6 TWcm$^{-2}$, along with Gaussian envelopes fitted to the individual fringe line-outs (dashed line), shown in the right panel.

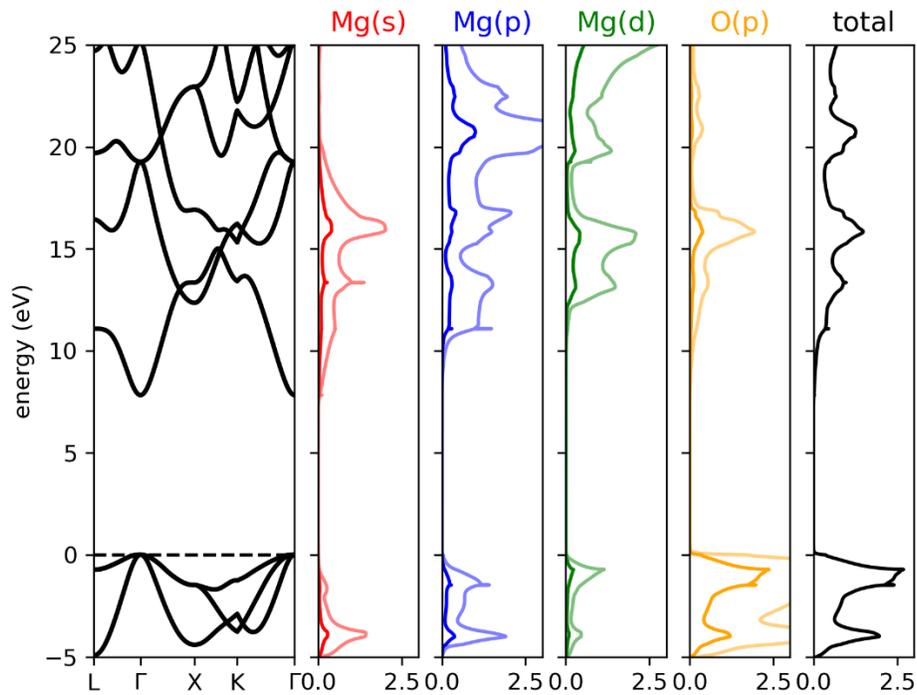

**Fig. S9 Orbital- and band-resolved multi-band numerical calculation results for MgO.**

The electronic band structure (left panel) and orbital-resolved projected density of states (PDOS) (right panels) in MgO. The zero-energy level is set to the VB maximum (dashed line). The direct band gap is 7.8 eV at the Γ-point. Four PDOS plots for the magnesium Mg (s, p, d) and oxygen O (p) orbitals are depicted in different colors, with the total PDOS represented in black. The two lines per panel are original data and a zoom (x5) thereof.

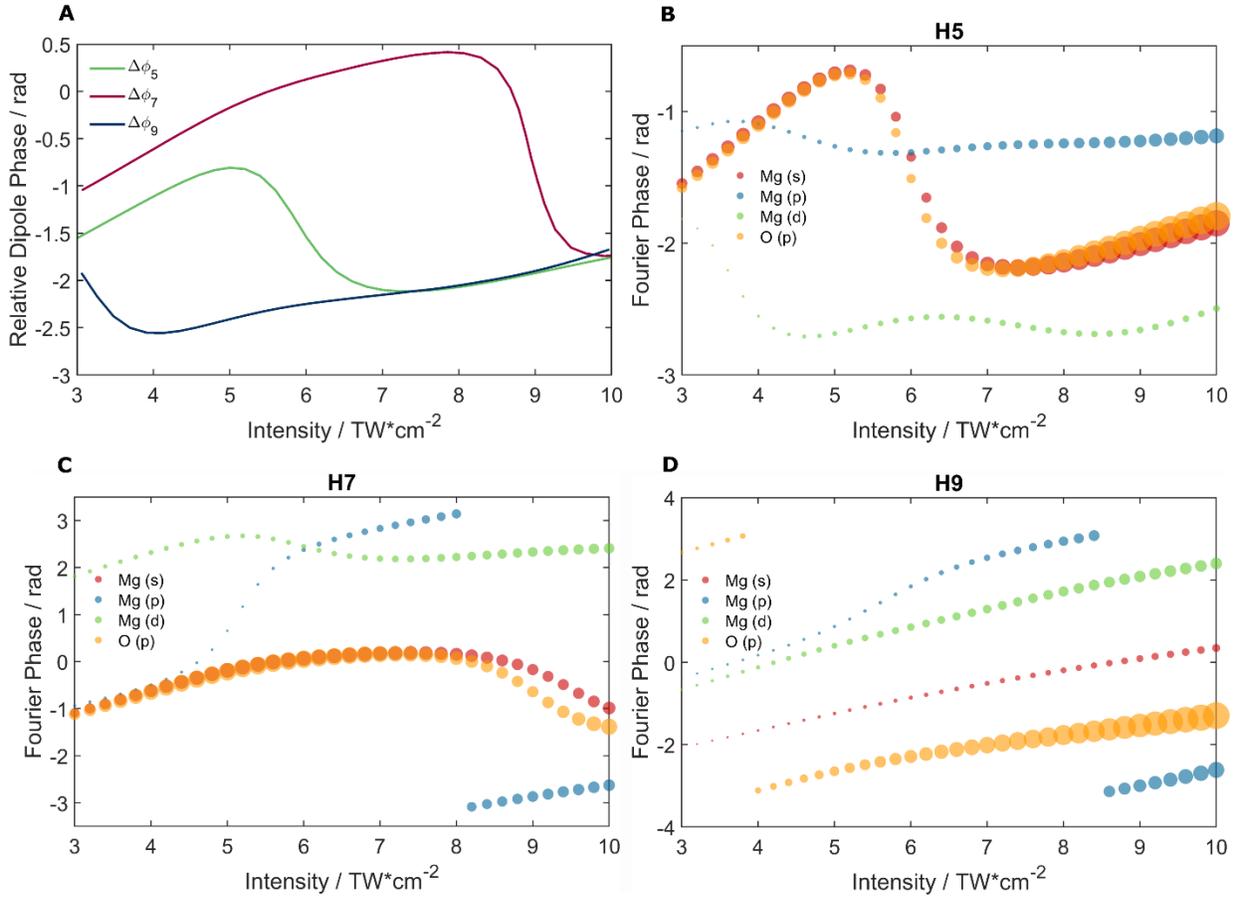

**Fig. S10 Multi-band numerical simulations for the Mg–Mg bond (Γ–K–X direction).**
(A) Simulated relative dipole phase results for harmonics 5 ($\Delta\varphi_5$), 7 ($\Delta\varphi_7$), and 9 ($\Delta\varphi_9$) in MgO with the 800 nm driving pulses aligned along the Mg–Mg bond (Γ–K–X direction), within the 3–10 TWcm$^{-2}$ intensity range, using the multi-band model. The Fourier phases and amplitudes of the Mg (s, p, d) and O (p) orbitals as a function of the 800 nm peak intensity, derived from the dipole phase simulations, are presented for the harmonics 5 (H5, 8 eV) (B), 7 (H7, 11 eV) (C), and 9 (H9, 14 eV) (D). The size of each circle indicates the magnitude of its Fourier amplitude, with distinct colors representing different orbitals.

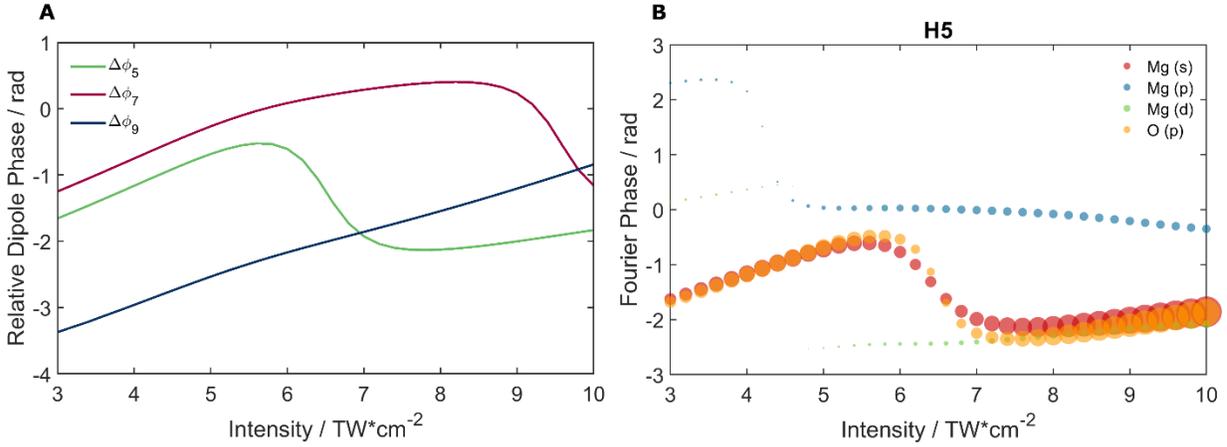

**Fig. S11 Multi-band numerical simulations for the Mg–O bond (Γ–X direction).**
(A) Simulated relative dipole phase results for harmonic 5 ($\Delta\varphi_5$), using the multi-band model, in comparison with those presented in the main text for harmonics 7 ($\Delta\varphi_7$) and 9 ($\Delta\varphi_9$) in MgO, along the Mg–O bond (Γ–X direction), within the 3–10 TWcm$^{-2}$ intensity range. The Fourier phases and amplitudes of the Mg (s, p, d) and O (p) orbitals for harmonic 5 (H5, 8 eV), as obtained from the orbital-based multi-band analysis (b). The size of each circle indicates the magnitude of its Fourier amplitude, with distinct colors representing different orbitals.

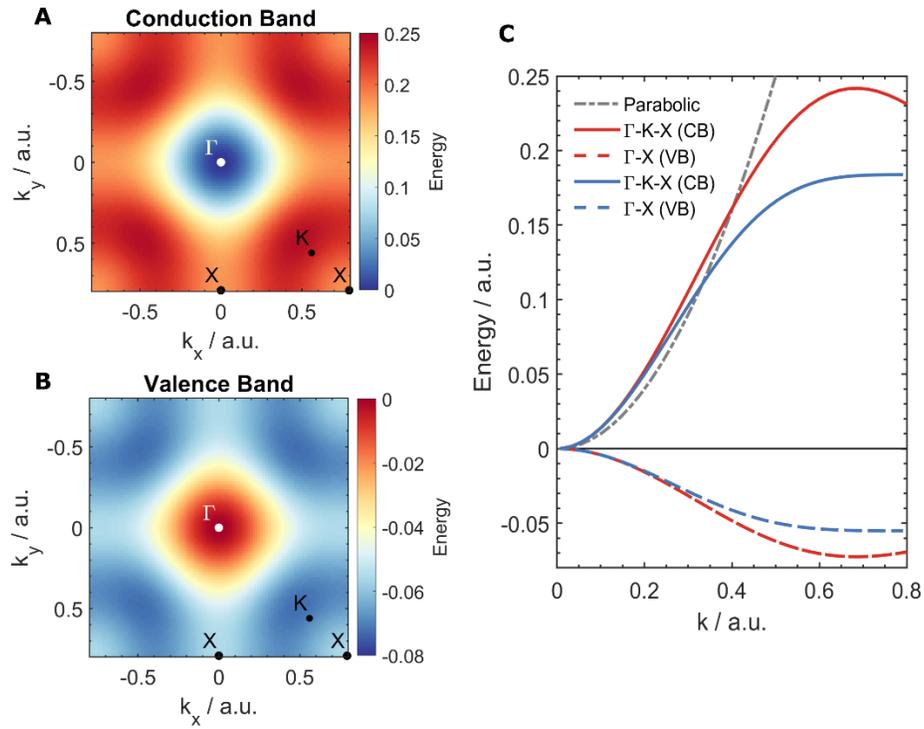

**Fig. S12 MgO crystal and band structure within two-band analytical and numerical model frameworks.**

The crystal structure of MgO (100) in a momentum space for the conduction band (CB) (A) and valence band (VB) (B). Panel (C) illustrates a cross-section of the bands along the Γ–X (Mg–O bond) and Γ–K–X (Mg–Mg bond) crystal directions, along with a reference parabola. For better visualization, the band gap has been set to zero.

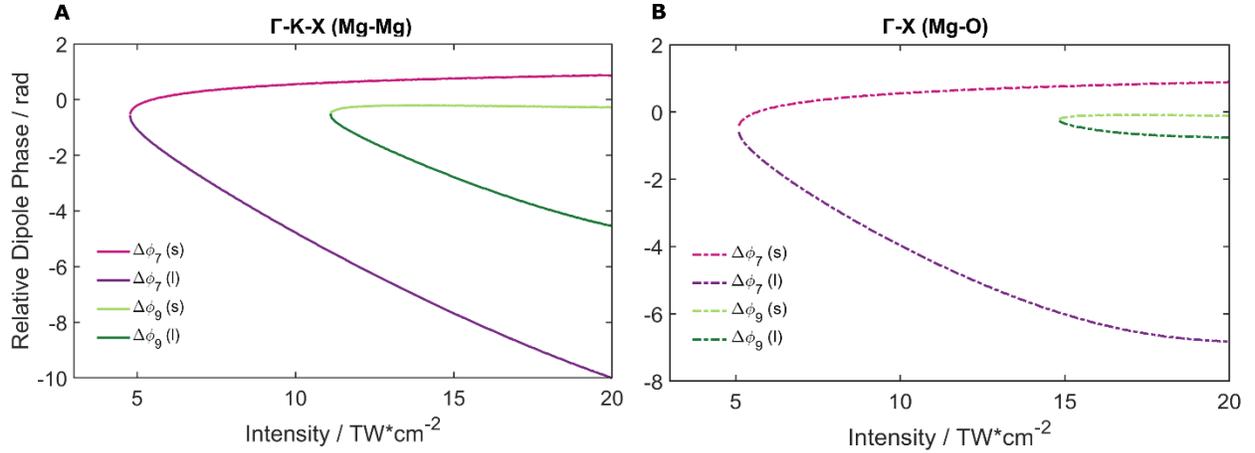

**Fig. S13 Two-band analytical simulations for the Γ–X / Γ–K–X directions.**

The calculated relative dipole phase as a function of the peak intensity with the 800 nm driver for harmonics 7 ($\Delta\varphi_7$), and 9 ($\Delta\varphi_9$), shown along the Γ–K–X (A) and Γ–X (B) crystal directions in MgO, derived from the two-band analytical semi-classical model. The figures highlight the intensity-dependent nature of dipole phase behavior for both short (s) and long (l) laser-driven electron trajectories, while also pinpointing the importance of the high-symmetry crystal direction involved in the generation of high-harmonic emission in MgO solid.